\documentclass[aps,prl,twocolumn,superscriptaddress]{revtex4}

\usepackage{graphicx}
\usepackage{epsfig}
\usepackage{here}

\begin{document}

\title{Erratum to: ``Nuclear Effects on $R = \sigma_L / \sigma_T$ in
                Deep-Inelastic Scattering" \\
Phys.~Lett.~B~475 (2000) 386
       }


\def\groupalberta{\affiliation{Department of Physics, University of Alberta, Edmonton, Alberta T6G 2J1, Canada}}
\def\groupargonne{\affiliation{Physics Division, Argonne National Laboratory, Argonne, Illinois 60439-4843, USA}}
\def\groupbari{\affiliation{Istituto Nazionale di Fisica Nucleare, Sezione di Bari, 70124 Bari, Italy}}
\def\groupcolorado{\affiliation{Nuclear Physics Laboratory, University of Colorado, Boulder, Colorado 80309-0446, USA}}
\def\groupdesy{\affiliation{DESY, Deutsches Elektronen-Synchrotron, 22603 Hamburg, Germany}}
\def\groupzeuthen{\affiliation{DESY Zeuthen, 15738 Zeuthen, Germany}}
\def\groupdubna{\affiliation{Joint Institute for Nuclear Research, 141980 Dubna, Russia}}
\def\grouperlangen{\affiliation{Physikalisches Institut, Universit\"at Erlangen-N\"urnberg, 91058 Erlangen, Germany}}
\def\groupferrara{\affiliation{Istituto Nazionale di Fisica Nucleare, Sezione di Ferrara and Dipartimento di Fisica, Universit\`a di Ferrara, 44100 Ferrara, Italy}}
\def\groupfrascati{\affiliation{Istituto Nazionale di Fisica Nucleare, Laboratori Nazionali di Frascati, 00044 Frascati, Italy}}
\def\groupfreiburg{\affiliation{Fakult\"at f\"ur Physik, Universit\"at Freiburg, 79104 Freiburg, Germany}}
\def\groupgent{\affiliation{Department of Subatomic and Radiation Physics, University of Gent, 9000 Gent, Belgium}}
\def\groupgiessen{\affiliation{Physikalisches Institut, Universit\"at Gie{\ss}en, 35392 Gie{\ss}en, Germany}}
\def\groupglasgow{\affiliation{Department of Physics and Astronomy, University of Glasgow, Glasgow G128 QQ, United Kingdom}}
\def\groupillinois{\affiliation{Department of Physics, University of Illinois, Urbana, Illinois 61801, USA}}
\def\groupliverpool{\affiliation{Physics Department, University of Liverpool, Liverpool L69 7ZE, United Kingdom}}
\def\groupwisconsin{\affiliation{Department of Physics, University of Wisconsin-Madison, Madison, Wisconsin 53706, USA}}
\def\groupmit{\affiliation{Laboratory for Nuclear Science, Massachusetts Institute of Technology, Cambridge, Massachusetts 02139, USA}}
\def\groupmichigan{\affiliation{Randall Laboratory of Physics, University of Michigan, Ann Arbor, Michigan 48109-1120, USA }}
\def\groupmoscow{\affiliation{Lebedev Physical Institute, 117924 Moscow, Russia}}
\def\groupmunich{\affiliation{Sektion Physik, Universit\"at M\"unchen, 85748 Garching, Germany}}
\def\groupnikhef{\affiliation{Nationaal Instituut voor Kernfysica en Hoge-Energiefysica (NIKHEF), 1009 DB Amsterdam, The Netherlands}}
\def\groupstpetersburg{\affiliation{Petersburg Nuclear Physics Institute, St. Petersburg, Gatchina, 188350 Russia}}
\def\groupprotvino{\affiliation{Institute for High Energy Physics, Protvino, Moscow oblast, 142284 Russia}}
\def\groupregensburg{\affiliation{Institut f\"ur Theoretische Physik, Universit\"at Regensburg, 93040 Regensburg, Germany}}
\def\grouprome{\affiliation{Istituto Nazionale di Fisica Nucleare, Sezione Roma 1, Gruppo Sanit\`a and Physics Laboratory, Istituto Superiore di Sanit\`a, 00161 Roma, Italy}}
\def\groupsimonfraser{\affiliation{Department of Physics, Simon Fraser University, Burnaby, British Columbia V5A 1S6, Canada}}
\def\grouptriumf{\affiliation{TRIUMF, Vancouver, British Columbia V6T 2A3, Canada}}
\def\grouptokyo{\affiliation{Department of Physics, Tokyo Institute of Technology, Tokyo 152, Japan}}
\def\groupamsterdam{\affiliation{Department of Physics and Astronomy, Vrije Universiteit, 1081 HV Amsterdam, The Netherlands}}
\def\groupyerevan{\affiliation{Yerevan Physics Institute, 375036 Yerevan, Armenia}}


\groupalberta
\groupargonne
\groupbari
\groupcolorado
\groupdesy
\groupzeuthen
\groupdubna
\grouperlangen
\groupferrara
\groupfrascati
\groupfreiburg
\groupgent
\groupgiessen
\groupglasgow
\groupillinois
\groupliverpool
\groupwisconsin
\groupmit
\groupmichigan
\groupmoscow
\groupmunich
\groupnikhef
\groupstpetersburg
\groupprotvino
\groupregensburg
\grouprome
\groupsimonfraser
\grouptriumf
\grouptokyo
\groupamsterdam
\groupyerevan


\author{A.~Airapetian}  \groupyerevan
\author{N.~Akopov}  \groupyerevan
\author{Z.~Akopov}  \groupyerevan
\author{M.~Amarian}  \grouprome \groupyerevan
\author{V.V.~Ammosov}  \groupprotvino
\author{E.C.~Aschenauer}  \groupzeuthen
\author{R.~Avakian}  \groupyerevan
\author{A.~Avetissian}  \groupyerevan
\author{E.~Avetissian}  \groupyerevan
\author{P.~Bailey}  \groupillinois
% ...

\collaboration{The HERMES Collaboration} \noaffiliation

\date{\today}

\begin{abstract}
\end{abstract}


\maketitle

\newpage
This erratum revokes the main conclusion of a Letter 
that reported measurements of cross sections for
deep-inelastic scattering (DIS) of leptons
on $^3$He and $^{14}$N targets, expressed as ratios of $\sigma_A / \sigma_D$ 
to the cross section on the
deuterium target~\cite{footnote1}.
In the particular kinematic domain 
$x < 0.03$ with $Q^2 < 1.25$ GeV$^2$,
$\sigma_A / \sigma_D$ was reported to differ as much as 35\% from earlier
such measurements at higher energies. As the only significant difference
from the earlier measurements appeared to be the kinematic variable
$y$, and hence the polarization parameter $\epsilon$, the new results
were interpreted as evidence for a nuclear influence on the
ratio $R$ of the cross sections for longitudinal and transverse photons.
This anomaly has now been discovered to be due to a peculiar
instrumental effect.

\begin{figure}[t]
\begin{center}
\includegraphics[width=7.5cm]{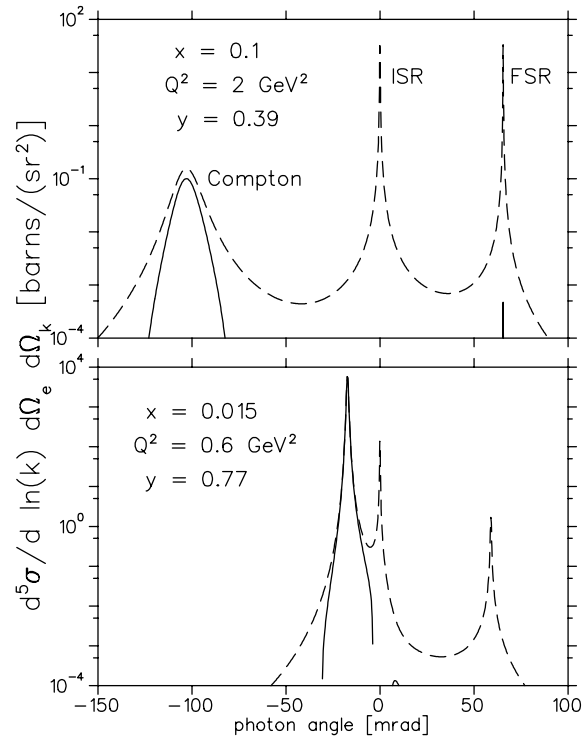}
\end{center}
\caption{The nuclear-elastic Bethe-Heitler cross 
section~\protect\cite{text} on $^{14}$N 
for two different 
coplanar kinematic conditions as labelled in terms of apparent DIS kinematic
variables. 
The continuous curves include the effects of the nuclear form factor. }
\label{fig:peaks}
\end{figure}

Radiative corrections do not cancel in the ratio $\sigma_A / \sigma_D$ 
because the yield of radiative events associated with elastic 
scattering scales nonlinearly, with $Z^2$.
At small values of {\em apparent} $x$ and $Q^2$ (inferred 
from the measured angle and energy of the scattered lepton), 
corresponding to large values of $y$, 
the contribution from
radiative elastic scattering becomes large. Unlike radiation associated
with inelastic processes, which is predominantly emitted 
in the direction of either the beam lepton (initial state radiation or ISR)
or the scattered lepton (final state radiation or FSR), the hard photons
associated with nuclear elastic scattering involve negligible
momentum transfer $q$ to the target nucleus (Compton peak). 
There are two reasons for this. One is that the Bethe-Heitler
cross section for radiative elastic processes predicts that in kinematic 
conditions corresponding to quite small values of apparent $x$ and $Q^2$, 
the Compton peak becomes much more prominant compared to ISR and FSR,
simply because smaller values of $q$ become kinematically available,
and the cross section is modulated by a factor of $1/q^4$.
This is illustrated in Fig.~\ref{fig:peaks}, which shows the nuclear-elastic
Bethe-Heitler cross section in two different coplanar kinematic situations,
both with and without including the nuclear form factor.
This latter comparison reveals the other reason --- that the nuclear form factor
strongly suppresses the cross section for significant
momentum transfer to the target, leaving only the Compton peak.

With negligible nuclear recoil momentum, 
essentially all of the transverse momentum
of the scattered lepton must be balanced by that of the radiated
hard photon, which also carries away most
of the beam energy at these large values of apparent $y$.
Hence we have
\begin{equation}
(1-y) \sin \theta_{e'} = y \sin \theta_\gamma,
\end{equation}
showing that at large $y$, the angle of the high-energy photon on 
the opposite side of the
beam line is correspondingly smaller than that of the scattered lepton,
but not negligible.
In the mirror-symmetric open geometry of the HERMES spectrometer,
this can have drastic consequences. These energetic photons
from nuclear targets have a high probability of hitting the 
detector frames surrounding the beam line 
in front of the dipole magnet, and producing extensive 
electromagetic showers that cause very high hit multiplicites in these
tracking detectors. For many of these nuclear-elastic events,
track reconstruction is therefore impossible, resulting in a large
tracking inefficiency that is strictly correlated with only this
process and kinematic situation. 

This problem is pernicious because
it is far from apparent in the experimental data. The event 
trigger rate for real DIS events is typically very small compared to
that from hadron background. Only after event reconstruction can all of the
particle identification criteria be applied to eliminate the hadrons.
However, event reconstruction is impossible for the affected radiative elastic
events, so they remain hidden in the dominant hadron background
and lost to the analysis, even though they are included in the radiative
corrections.
A simulation of the experiment reveals the problem only if it includes 
both the nuclear target with its particular radiative effects, 
and a complete treatment of showers 
in material {\em outside} of the geometric acceptance. This was not 
included in the data analysis for the original Letter. Subsequent such 
studies~\cite{newpaper} revealed that these showers can indeed account
for the differences between the HERMES results and previous measurements
in similar kinematic regions. The strong correlation
of the shower phenomenon with the kinematic variable $y$ accounts
for its correlation also with the photon polarization parameter $\epsilon$,
which led to the incorrect interpretation in terms of $R_A/R_D$. 
This interpretation became untenable when analyzing
$^{84}$Kr/$^2$H cross section ratios extracted from data collected after
the publication of the $^{14}$N/$^2$H data: 
for this heavy target nucleus the instrumental effects described above 
are so severe that the extracted Born cross section ratio became unphysical.
The corrected 
$^3$He/$^2$H and $^{14}$N/$^2$H data
and the new $^{84}$Kr/$^2$H data are presented in the 
accompanying paper~\cite{newpaper}.

The part of the HERMES kinematic region affected by the correlated background 
from nuclear targets is restricted to $x<0.06$ with $Q^2<2\,$GeV$^2$. 
Polarized DIS data from hydrogen, deuterium and helium-3 targets are
unaffected by this effect, because of both the more restricted
kinematic range, and the much smaller value of $Z^2$ modulating
the elastic Bethe-Heitler cross section. Semi-inclusive data are
also unaffected even with nuclear targets~\cite{seminuc}, as radiative elastic events 
are excluded by the presence of a hadron in the final state.


\begin{thebibliography}{}
\bibitem[*]{footnote1}{All symbols used here are defined in the
original Letter.}
\end{thebibliography}
\end{document}